\documentclass[11pt,a4paper]{article}

\let\ssection=\section
\renewcommand{\section}{\setcounter{equation}{0}\ssection}

\title{Symmetries of fluid dynamics\\
with polytropic exponent\footnote{Dedicated to the memory of
Lochlainn O'Raifeartaigh, our late friend and teacher.}
\\
}
\author{M.~Hassa\"{\i}ne and P. A.~Horv\'athy\\
\\
Institute for Theoretical Physics, Roland E\"{o}tv\"{o}s
University,\\
H-1117, BUDAPEST, P\'azm\'any P. s\'et\'any 1/A (Hungary)
\\
and
\\
Laboratoire de Math\'ematiques et de Physique Th\'eorique\\
Universit\'e de Tours, Parc de Grandmont\\
F-37200 TOURS (France)\\
}
\def\D{{\cal D}}
\def\K{{\cal K}}
\def\H{{\cal H}}
\def\IR{{\bf R}}
\def\smallcirc{{\raise 0.5pt \hbox{$\scriptstyle\circ$}}}
\def\smallover#1/#2{\hbox{$\textstyle{#1\over#2}$}} %

\def\parag{\hfil\break} 
\def\kikezd{\parag\underbar}

\begin{document}

\maketitle
\begin{abstract}
The symmetries of the general Euler equations of fluid dynamics with
polytropic exponent are determined using the Kaluza-Klein
type framework of Duval et $\it{al}$.
In the standard polytropic case the recent results of O'Raifeartaigh
and Sreedhar are confirmed. Similar
results are proved for
 polytropic exponent $\gamma=-1$, which corresponds to
the dimensional reduction of $d$-branes.
The relation between the duality transformation used in describing
supernova explosion and Cosmology is explained.
\end{abstract}

\section{Introduction}

The amazing similarity of supernova explosion and plasma implosion
has been explained not less amazingly by Drury and
Mendon\c ca \cite{DM}, who pointed out that the two situations
 can be related by the ``duality'' transformation
$
\Sigma~: t\to -1/t,\; {\bf x}\to {\bf x}/t$.
This strange-looking transformation
belongs to the ${\rm SL}(2,\IR)$  group generated by
the
dilatations, $\D:t\to \delta^2t,\, {\bf x}\to \delta{\bf x}$,
expansions,
$\K~: t\to t/1+\kappa t,\; {\bf x}\to{\bf x}(1+\kappa t)^{-1}$,
 and time-translation, ${\cal H}~:t\to t+\epsilon$,
which are indeed symmetries of a free non-relativistic particle
\cite{NH, DGH}. In fact, $\Sigma=\H_{-1}\smallcirc \K_{1}\smallcirc
H_{-1}$.

Motivated by the results of Drury and Mendon\c ca, O'Raifeartaigh
and Sreedhar \cite{RS}
performed a systematic study of the symmetries of the Euler equations
of fluid dynamics,
\begin{eqnarray}
 D\rho & = &-\rho\vec{\nabla}\cdot{\bf u},\\
\rho\, D{\bf u} & = &-\Lambda(\gamma-1)\vec{\nabla}(\chi
\rho^{\gamma})+{\bf V},\\
D\chi & = &0,
\label{eulergeneral}
\end{eqnarray}
where $D$ is the convective derivative,
$D=\partial_t+{\bf u}\cdot\vec{\nabla}$, and the fields $\rho$ and
${\bf u}$ are the density and the velocity. ${\bf V}$ is
the viscosity term, with components
\begin{eqnarray}
V_i=\partial_j\left(\eta\left(\partial_j u_i+\partial_iu_j
-\frac{2}{d}\delta_{ij}\partial_k u_k\right)\right)+\partial_i
\left(\xi\partial_k u_k\right),
\label{viscosityterms}
\end{eqnarray}
where  $d$ is the spatial dimension,
$\xi$ and $\eta$ represent the bulk  and shear viscosity fields,
respectively.  $\gamma$ is the polytropic exponent and
 $\Lambda$ is the coupling constant of a
potential, $U(\rho)=\Lambda\, \rho^{\gamma}$. The field $\chi$ is
related to the energy density $\epsilon$ by $\epsilon=\chi\rho^{\gamma}$.

O'Raifeartaigh and Sreedhar consider first the sub-class of
($1.1$)-($1.3$) characterised
by
$(i)$ the absence of viscosity terms, ${\bf V}=0$;
$(ii)$ the dynamical field $\chi$ is choosen to be
$\chi=1$;
$(iii)$ the motion is assumed irrotational,
${\rm rot}\,{\bf u}=0$.
Then they show that
when the polytropic exponent takes the standard value $\gamma=1+2/d$,
the equations
($1.1$)-($1.3$) are invariant w.r.t. Schr\"{o}dinger
transformations, composed of Galilei transformations, augmented by
dilatations and expansions
\cite{NH}. When the conditions $(i), (ii)$ and $(iii)$ are relaxed,
the expansions are generally broken by the viscosity term;
dilatations remain, however, symmetries \cite{RS}.

Similar questions were
investigated by Bordemann and Hoppe, and Jevicki \cite{BoHo},
and by Jackiw, Polychronakos, and Bazeia
\cite{JAC, BJ}, who found that the dimensional reduction
of d-brane theory yields a
viscosity--free, isentropic and irrotational hydrodynamical model
called the Chaplygin gas,
eqns. ($1.1$)-($1.3$) with ${\bf V}=0$ and $\chi=1$ and
with effective
potential $U \propto 1/\rho$.
Remarkably, their system
admits a hidden Poincar\'e symmetry \cite{BoHo, JAC, BJ},
composed of the Galilei transformations,
augmented by $(d+1)$ generators we called time-dilatation and antiboost
 \cite{HH}.

\goodbreak
In this Letter, we combine and generalize these results in a unified
framework.
First, we confirm  the results of O'Raifeartaigh
et {\it al}. by dropping condition $(iii)$ right on from the
beginning.
Then we extend the d-brane results in \cite{BoHo, JAC, BJ}
showing that, for $U \propto 1/\rho$, the symmetries of
the general equations ($1.1$)-($1.3$) with conditions $(i)$ and $(ii)$ alone
 still admit a Poincar\'e symmetry.
Viscosity breaks part of this large symmetry.
There remains, however,  time-dilatation,
$\Delta~: t\to e^{\alpha}t,\; {\bf x}\to {\bf x}$, analogous to
dilatations, $\D$,  in the standard case.
\goodbreak

 The relation of the duality transformation
$\Sigma$ and newtonian cosmology is also explained.
Although our results could also be obtained in a classical
approach \cite{NH,  RS, BJ},
we found it more convenient to use  Duval's Kaluza
Klein--type framework \cite{DGH}, which sheds a new light
on the arisal of these symmetries.

\section{Symmetries of the Euler equations}

The simplest way to confirm the result of O'Raifeartaigh and
Sreedhar \cite{RS},
is to consider \cite{HH}, Sect. 2, p. 224
(see also \cite{JPRIV}), the stress--energy tensor $T^{\alpha\beta}$.
In the absence of viscosity, ${\bf V}=0$
and for $\chi=1$, they are given, e. g., in Eq. (2.2) in the first
reference of
\cite{JAC}, as
\begin{equation}
	T^{00}=\rho\frac{{\bf u}^2}{2}+U(\rho),
	\qquad
	T^{ij}=\rho u^{i}u^{j}-\delta^{ij}(U-\rho\partial_{\rho}U),
    \label{emtensor}
\end{equation}
where $\partial_{\rho}U$ is the enthalpy\footnote{It is worth
noting that, although it has been derived assuming irrotationality,
(\ref{emtensor}) actually provides us with a conserved energy-momentum
tensor
in the general case, as it can be verified by a directly,
using the Euler equations.}.
Next recall (e. g.
\cite{JP}, Eq. (2.261) ) the criterion
of Schr\"odinger symmetry:
\begin{equation}
    2T^{00}=\sum_{i}T^{ii},
    \label{nrtracecond}
 \end{equation}
which replaces, in the non-relativistic context,
the familiar  condition for relativistic conformal
invariance, {\it viz}. $T^{\mu}_{\ \mu}=0$. With the above expression
for $T^{00}$ and $T^{ij}$, we get
 a differential equation for $U$, namely
$\rho\partial_{\rho}U=(2/d+1)U$ or
$
U=\Lambda\rho^{1+2/d},
$
which is the result in \cite{RS}.

More generally, let us first consider
the sub-class of ($1.1$)-($1.3$) with
conditions $(i)$ and $(ii)$ alone. Using the
Clebsch parametrization \cite{CLE},
${\bf u}=\vec{\nabla}\phi-\nu\vec{\nabla}\theta$,
provides us with a local lagrangian theory \cite{RS}.
Then eliminating the Lagrange multiplier $\nu$
yields the equations of motion
\begin{eqnarray}
\begin{array}{ll}
\left(E_{\gamma}\right) &
\left\lbrace
\begin{array}{l}
\partial_t\rho+\partial_k\left(\rho\partial_k\phi+
\displaystyle{\frac{\rho}
{\vert\vec{\nabla}\theta\vert^2}}\partial_k
\theta\left(\partial_t\theta+
{\vec\nabla}
\theta\cdot\vec{\nabla}\phi\right)\right)=0,
\\[3mm]
\partial_t\phi+\frac{1}{2}
\vert{\vec\nabla}\phi\vert^2+
\displaystyle{\frac{1}
{2\vert{\vec\nabla}\theta\vert^2}}
\left(\partial_t\theta+
\vec{\nabla}\theta\cdot{\vec\nabla}\phi\right)^2
-\gamma\Lambda\rho^{\gamma-1}=0,
\\[3mm]
\partial_t\left(\displaystyle{\frac{\rho}{\vert{\bf
\vec\nabla}\theta\vert^2}}\left(\partial_t\theta
+\vec{\nabla}\theta\cdot{\vec\nabla}\phi\right)\right)+
\\[2mm]
\qquad
\partial_k\left(\displaystyle{\frac{\rho\,\partial_k\phi}
{\vert{{\vec{\nabla}}}\theta\vert^2}}\left(\partial_t\theta
+\vec{\nabla}\theta\cdot{{\vec{\nabla}}}\phi\right)
-\displaystyle{\frac{\rho\,\partial_k\theta}
{\vert{{\vec{\nabla}}}\theta\vert^{4}}}
\left(\partial_t\theta
+\vec{\nabla}
\theta\cdot{{\vec{\nabla}}}\phi\right)^2\right)=0.
\label{equationsQ}
\end{array}
\right.
\end{array}
\end{eqnarray}
The velocity field
${\bf u}$ here is expressed in terms of $\theta$ and $\phi$ by
${\bf u}=\vec{\nabla}\phi-
(\vec{\nabla}\theta/\vert{\vec{\nabla}}\theta\vert^2)
\left(\partial_t\theta
+\vec{\nabla}\theta\cdot{\vec\nabla}\phi\right).
$

Below we analyse the symmetries of
(\ref{equationsQ}) in the Kaluza-Klein type framework of \cite{DGH}.
Non-relativistic
space-time, $Q$, has coordinates $({\bf x},t)$, and
can also be obtained from one higher dimensional manifold
$M$ with coordinates $({\bf x},t,s)$, when the coordinate $s$
 is factored out.
$M$ is endowed with the flat Lorentz
metric $d{\bf x}^2+2dtds$;
$\Xi=\partial_s$ light-like vector field.
$M$ is a relativistic spacetime, upon which we consider
the real fields $R$, $\Theta$ and $\Phi$. Inspired by (\ref{equationsQ}),
we postulate
\begin{eqnarray}
\begin{array}{ll}
\left({\cal E}_{\gamma}\right) &
\left\lbrace
\begin{array}{l}
\partial_\mu\left(\displaystyle{\frac{R}{2}}
\,\partial^\mu\Phi+\displaystyle
{\frac{R\,\partial^\mu\Theta}
{(\partial_\sigma\Theta)(\partial^\sigma\Theta)}}
\partial_\nu\Theta\,\partial^\nu\Phi\right)=0
\\[3,5mm]
\partial_\mu\Phi\,\partial^\mu\Phi
+\displaystyle{\frac{1}{(\partial_\mu\Theta)
(\partial^\mu\Theta)}}\left(\partial_\nu\Phi\,
\partial^\nu\Theta\right)^2-
\gamma\Lambda
R^{\gamma-1}=0,\\[3,5mm]
\partial_\mu\left(R\,\partial^\mu\Phi\,
\displaystyle{\frac{\partial_\sigma\Phi\,
\partial^\sigma\Theta}
{(\partial_\nu\Theta\,\partial^\nu\Theta)}}-\frac{R}{2}\,
\partial^\mu\Theta\,
\displaystyle{\frac{(\partial_\sigma\Phi\,\partial^\sigma
\Theta)^2}
{(\partial_\nu\Theta\,\partial^\nu\Theta)^2}}\right)=0.
\label{equationsM}
\end{array}
\right.
\end{array}
\end{eqnarray}

To complete our Kaluza-Klein framework, we need to establish a correspondance
between the systems $(\ref{equationsQ})$ and $(\ref{equationsM})$.
Below we define, for both critical values of $\gamma$, a judicious (and
different) relation between the fields on $M$ and those on $Q$, such
that the relativistic system $\displaystyle{({\cal E}_{\gamma})}$
projects to the non-relativistic one
$\displaystyle{({E}_{\gamma})}$. Then the
symmetries of the latter arise by projection.

$\bullet$ Let us first consider the standard case, $\gamma=1+2/d$.
If the fields $R$,
$\Theta$ and $\Phi$
are of the particular form
\begin{eqnarray}
R({\bf x},t,s)=
\rho({\bf x},t),\;\;
\Theta({\bf x},t,s)=\theta({\bf x},t)\;\;
\Phi({\bf x},t,s)=\phi({\bf x},t)+s
\label{equivarianceduval}
\end{eqnarray}
(which is in fact the usual equivariance condition \cite{DGH}),
then the equations $\displaystyle{({\cal
E}_{1+2/d})}$ project to
$\displaystyle{({E}_{1+2/d})}$.

Now we determine the symmetries. One shows readily  that
 if the fields $R$,
$\Phi$ and $\Theta$ are solutions of equations
$\displaystyle{({\cal E}_{1+2/d})}$,
then their images under a conformal
transformation of $M$, $\varphi^\star g=\Omega^2 g$, implemented as
$\tilde{R}=\Omega^d\,\varphi^\star R$,
$\tilde{\Phi}=\varphi^\star\Phi$ and
$\tilde{\Theta}=\,\varphi^\star\Theta$, also satisfy the same equations.
They are hence symmetries for (\ref{equationsM}).
To make the transformed
fields equivariant in the sense (\ref{equivarianceduval}),
however, we must restrict
ourselves to transformations
which preserve the ``vertical''vector field $\Xi$. Their action on
$M$,
\begin{eqnarray}
\left\lbrace
\begin{array}{l}
\tilde{{\bf x}}={\vec\gamma}-{\vec\beta}t+
\displaystyle{\frac{\delta\,{\cal
R}{\bf x}}{(1+\kappa t)}},
\\[2mm]
\tilde{t}=\displaystyle{\frac{\epsilon+\delta^2\,t}{(1+\kappa t)}},
\\[2mm]
\tilde{s}=s+\lambda(t,{\bf x}),
\qquad
\lambda(t,{\bf x})\equiv
{\vec\beta}\cdot{\bf x}-\frac{1}{2}\vert{\vec\beta}\vert^2\,t
+\displaystyle{\frac{\kappa}{2}
\frac{\vert{\bf x}\vert^2}{(1+\kappa t)}},
\label{transfM}
\end{array}
\right.
\end{eqnarray}
(where ${\cal R}\in so(2)$, ${\vec\beta},
{\vec\gamma}, \epsilon, \kappa$ and
$\delta$ are interpreted as rotation, boost, space
translation, time translation, expansion and dilatation)
projects into non-relativistic space-time, $Q$, according to the classical
Schr\"{o}dinger transformations
\cite{NH, DGH}. The action on fields are
obtained by using the previous relations. Setting
$M=\displaystyle{\left(\partial \tilde{x}_i/\partial x_j\right)}$, we
get
\begin{eqnarray}
\begin{array}{l}
\left\lbrace
\begin{array}{l}
\tilde{\rho}(t,{\bf x})=\displaystyle{\frac{\delta^d}{(1+\kappa
t)^d}}\,\rho(\tilde{t},\tilde{{\bf
x}})=\hbox{det}(M)\,\rho(\tilde{t},\tilde{{\bf x}}),
\\[2.8mm]
\tilde{\phi}(t,{\bf x})=\phi(\tilde{t},\tilde{{\bf x}})
+\lambda(t,{\bf x}),
\\[2mm]
\tilde{\theta}(t,{\bf x})=
\theta(\tilde{t},\tilde{{\bf x}}).
\end{array}
\right.
\end{array}
\label{dilatation}
\end{eqnarray}
\goodbreak

Since the $\Xi$-preserving symmetries of (\ref{equationsM}) project to
symmetries, we
 conclude that, in the viscosity--free case $\xi=\eta=0$, the
(not necessarily irrotational)
system has a full Schr\"{o}dinger symmetry, as stated above.

Another way of reaching this result, closer in spirit to our first proof,
is to observe that
 Eqns. (\ref{equationsM}) derive from the relativistic Action
\begin{equation}
S=\int\Big(R\partial_\mu\Phi\,\partial^\mu\Phi
+\frac{R}{\partial_\mu\Theta\,\partial^\mu\Theta}\,
\left(\partial_\sigma\Phi\,
\partial^\sigma\Theta\right)^2-2\Lambda
R^{\gamma}\Big)\sqrt{g}d^{d+2}x,
\end{equation}
where, for convenience, we moved to a general Lorentz metric
$g_{\mu\nu}$ on $M$.
The associated energy-momentum tensor
${\cal T}_{\mu\nu}=2\delta S/\delta g^{\mu\nu}$, i.~e.,
\begin{eqnarray}
{\cal T}_{\mu\nu} &=& R\,\partial_\mu\Phi\,\partial_\nu\Phi
-\frac{R}{2}(\partial_\sigma\Phi\,\partial^\sigma\Phi)\,g_{\mu\nu}
+\Lambda\,R^{\gamma}g_{\mu\nu}\nonumber\\
     &+&\frac{R}{\partial_\sigma\Theta\,\partial^\sigma\Theta}
     \left(\partial_\mu\Phi\,\partial_\nu\Theta
     +\partial_\mu\Theta\,\partial_\nu\Phi\right)
     (\partial_\sigma\Theta\,\partial^\sigma\Phi)\nonumber\\
      &-&R\,
      \partial_\mu\Theta\,\partial_\nu\Theta
      \frac{\left(\partial_\sigma\Theta\,\partial^\sigma\Phi\right)^2}
{\left(\partial_\sigma\Theta\,\partial^\sigma\Theta\right)^2}
	   -\frac{R}{2}g_{\mu\nu}
	   \frac{\left(\partial_\sigma\Theta\,\partial^\sigma\Phi
	   \right)^2}
{\left(\partial_\sigma\Theta\,\partial^\sigma\Theta\right)},
\end{eqnarray}
(which generalizes the expression given in \cite{HH})
is seen to be symmetric and conserved. Relativistic conformal
invariance  requires the vanishing of its trace,
\begin{equation}
\sum_{\mu}{\cal T}^\mu_{\ \mu}=
\Lambda d\,R^{\gamma} \left(\gamma-\big[1+
\frac{2}{d}]\right)=0,
\label{relattrace}
\end{equation}
which yields the correct polytropic exponent
$\gamma=1+2/d$ once again. To conclude, the Schr\"odinger
group is the $\Xi$-preserving part of the (relativistic) conformal group.
It is worth mentionning that the $ti,\, it$ and $ij$ components of
 the relativistic ${\cal T}^{\mu\nu}$ are related to
 the non-relativistic $T^{\alpha\beta}$ by surface terms,
 and that the non-relativistic trace condition (\ref{nrtracecond})
 follows from $-T^{00}={\cal T}^s_{\ s}={\cal T}^t_{\ t}$.

\goodbreak
Let us now return to the general equations ($1.1$)-($1.3$)
 including viscosity.  We first determine  how ${\bf u}$
transforms. Let us define on $M$ an $s$-independent  vector
$(k_\nu)\equiv (k_t,{\bf u},k_s)$,
\begin{eqnarray}
k_\nu=
\partial_\nu\Phi-\displaystyle{\frac{\partial_\nu\Theta}
{\left(\partial_
\sigma\Theta
\,\partial^\sigma\Theta\right)}\left(\partial_\mu\Theta
\,\partial^\mu\Phi\right)}.
\label{budapest}
\end{eqnarray}
Using the transformation rule on $M$ of this vector,
$\displaystyle{\tilde{k}_\mu=
\left(\partial \tilde{x}^\nu/\partial {x}^\mu
\right)k_\nu}$, the action on ${\bf u}$, the
space component of $k_{\nu}$, is obtained, namely
\begin{eqnarray}
\tilde{{\bf u}}\,(t,{\bf x})=[{\cal
R}\,(\hbox{det}\,M)^{1/d}]\,{\bf u}\,(\tilde{t},{\tilde{{\bf x}}})
+\vec{\nabla}\lambda.
\end{eqnarray}

It is interesting to observe that the restriction (ii), {\it viz}.
$\chi=1$, can actually be relaxed:
 the viscosity--free Euler equations
 are invariant w.r.t. transformations (\ref{transfM})
and (\ref{dilatation}), whenever  $\tilde{\chi}=\chi$.
The first term in  (1.2), $\rho D{\bf u}$,  transforms in fact into
$({\rm det}\,M)^{1+3/d}\rho D{\bf u}$, and if
$\tilde{\chi}=\chi$, then the term $\vec{\nabla}(\chi\rho^{1+2/d})$ becomes
$({\rm det}\,M)^{1+3/d}\vec{\nabla}(\chi\rho^{1+2/d})$ so that eqn. (1.2)
merely gets multiplied by an overall factor. The other equations are
plainly invariant.

Now, if
$\tilde{\eta}=(\hbox{det}\,M)\,\eta$ and
$\tilde{\xi}=(\hbox{det}\,M)\,\xi$, the viscosity term
transforms as
\begin{equation}
   \begin{array}{c}
V_i\to \tilde{V}_i=(\hbox{det}\,M)^{1+\frac{3}{d}}\,V_i+\hfill
\\[2mm]
(\hbox{det}\,M)^{1+\frac{1}{d}}
\left[\tilde{\partial}_i(\xi\Delta
\lambda)+\tilde{\partial}_j\left(\eta[2
\partial_i\partial_j\lambda-\frac{2}{d}\delta_{ij}\Delta
\lambda]\right)\right].
\label{obstruct}
\end{array}
\end{equation}

Invariance of Eqn. $(1.2)$ requires the second term here to vanish.
For  $\lambda$ in (\ref{transfM}), this is automatical for the shear
viscosity field $\eta$. The bulk viscosity field, $\xi$, however,
breaks the expansions, leaving us with dilatational
symmetry only. For time-independent fields
one also have time-translations. (This is consistent, owing to $\{\H,
\D\}=\H$). When the viscosity fields only depend on time, though,
the residual symmetry includes the expansions but break the
time-translational invariance. These results confirm the conclusion
of \cite{RS} obtained in a rather different way.

$\bullet$ Next, we consider the d-brane potential, $\gamma=-1$. The
``non-relativistic conformal symmetries'' (i. e. dilatations and
expansions) are plainly broken.
However, when the motion is irrotional and viscosity--free, this
$(d+1)$ dimensional non-relativistic model
 admits the $(d+1,1)$-dimensional Poincar\'e group
 as symmetry \cite{BoHo, JAC, BJ}. Generalising the results and
 the procedure presented in
\cite{HH}, now we show that the not necessarily irrotational but
still viscosity--free system
$\displaystyle{({E}_{-1})}$ is Poincar\'e
symmetric. Our previous equivariance condition (\ref{equivarianceduval})
is seen to be be too restrictive and we propose to
 relate  instead
the fields defined on $M$ and $Q$
according to
\begin{eqnarray}
\left\lbrace
\begin{array}{l}
\rho({\bf x},t)=R\left({\bf x},t,-\phi({\bf
x},t)\right)\,\partial_s\Phi\left({\bf x},t,-\phi({\bf
x},t)\right),\\[1,2mm]
\Phi\left({\bf x},t,-\phi({\bf x},t)\right)=0,\\[1.2mm]
\Theta({\bf x},t,s)=\theta({\bf x},t).
\end{array}
\right.
\label{equivariancehh}
\end{eqnarray}

Here the point
$\left(t,{\bf x},-\phi(t,{\bf x})\right)$ in $M$ is defined as a zero
of the field $\Phi=0$. Note that $R$
can depend on the $s$ variable; however,
 $\rho$ is already defined  $Q$.
It is easy to see
that this condition is more general than  classical equivariance
(\ref{equivarianceduval}). As previously,
$\displaystyle{({\cal E}_{-1})}$
with the constraint (\ref{equivariancehh}), project into $Q$ as
$\displaystyle{({E}_{-1})}$. Let us  insist that this projection is
only possible for the d-brane potential \cite{HH}. The advantage of the general
equivariance is that, now, we can consider
transformations which do not  necessarily preserve $\Xi$.
But the
particular form of our potential restricts ourselves to
consider only isometric transformations. These latter are symmetries of
equations
$\displaystyle{({\cal E}_{-1})}$ coupled to the constraint
(\ref{equivariancehh}). The action of the $\Xi$-preserving
isometries lead to the extended Galilei transformations. The
non-preserving part is composed by $(d+1)$ generators whose action on $M$
is given by \cite{HH} :
\begin{eqnarray}
\begin{array}{l}
\left\lbrace
\begin{array}{l}
\tilde{{\bf x}}={\bf x}-\vec{\omega}s
\\
\tilde{t}=
e^{\alpha}\left(t+\vec{\omega}\cdot{\bf x}-\frac{1}{2}
\vert\vec{\omega}\vert^2\,s\right),
\\
\tilde{s}=e^{-\alpha}s,
\end{array}
\right.
\end{array}
\label{timedilatationM}
\end{eqnarray}
where $\alpha$ and $\vec{\omega}$ are the parameters associated with
time dilatation and antiboost, respectively.
Our transformations act on
fields naturally, as
$\tilde{R}(x,t,s)=R(\tilde{\bf x},\tilde{t},\tilde{s})$, etc.
The projection into $Q$ yields  \cite{JAC, BJ}
\begin{eqnarray}
\begin{array}{lllll}
\left\lbrace
\begin{array}{l}
\tilde{\bf x}={\bf x}+\vec{\omega}\,\phi(\tilde{\bf x},\tilde{t}),

\\[2mm]
\tilde{t}=e^{\alpha}
\left(t+\frac{1}{2}\vec{\omega}\cdot({\bf x}+
\tilde{\bf x})\right)
\end{array}
\right.
&
\hbox{and}
&
\left\lbrace
\begin{array}{l}
\tilde{\rho}({\bf x},t)=\rho(\tilde{\bf x},\tilde{t})\,J^{-1}
\\[2mm]
\tilde{\phi}({\bf x},t)=e^\alpha\phi(\tilde{\bf x},\tilde{t})
\\[2mm]
\tilde{\theta}({\bf x},t)=\theta(\tilde{\bf x},\tilde{t})
\end{array}
\right.
&
\hbox{}
\end{array}
\label{lise}
\end{eqnarray}
where $J$ is the Jacobian of the transformation given by
\begin{eqnarray}
J=e^\alpha\left[1-\sum_k\,\omega_k\,\tilde{\partial}_{k}
\phi(\tilde{\bf x},\tilde{t})
-\frac{1}{2}\vert\vec{\omega}\vert^2\,\partial_{\tilde{t}}
\phi(\tilde{\bf x},\tilde{t})\right]^{-1}.
\end{eqnarray}

As in the standard case, the vector $k_\mu$
(\ref{budapest}) can be used to determine the transformation on the
velocity. But now because of this particular equivariance,
the velocity is equal to
${\bf u}=\displaystyle{({\bf k}/\partial_s\Phi)(t,{\bf x},
-\phi(t,{\bf x}))}$ and a similar calculation yields instead
\begin{eqnarray}
\tilde{{\bf u}}(t,{\bf x})=J\left[{\bf u}(\tilde{t},\tilde{\bf x})
+\vec{\omega}
\left(\tilde{\partial}_{t}\phi-
\frac{\partial_{\tilde{t}}\theta}{\sum(
\tilde{\partial}_{k}\theta)^2}
\left(\partial_{\tilde{t}}\theta+\tilde{\partial}_{m}
\theta\;\tilde{\partial}_{m}\phi\right)
\right)\right].
\label{chat}
\end{eqnarray}

As in
the standard case, the viscosity term breaks most of the symmetry. A rather
tedious
calculation shows in fact that, under a Poincar\'e transformation,
the viscosity term (\ref{viscosityterms})
transforms as
\begin{equation}
    \tilde{V}_{i}=e^{\alpha}\,V_{i}+F(\vec{\omega},\xi,\eta),
\end{equation}
where $F(\vec{\omega},\xi,\eta)$ is a complicated expression
which vanishes for $\vec{\omega},\ \xi,$ or $\eta$ equal zero.
For non-trivial viscosity, this means that the antiboosts are broken.
Eq. (1.2) is, however, merely multiplied by $e^{\alpha}$
 under $\Delta~: t\to e^{\alpha}t$~:
time  (rather then
non-relativistic) dilatation, $\Delta$,
is a residual symmetry.

\section{Explosion/implosion duality and cosmology}

The clue of Drury and Mendon\c ca \cite{DM} is to map, using the ``duality
transformation''
 $\Sigma: \tilde{t}=-1/t,\;
\tilde{\bf x}={\bf x}/t$, supernova
{\it explosion} at time $t=0$
into  an {\it implosion} starting at $\tilde{t}=-\infty$ and evolving to
$\tilde{t}=0$. Then they find that, implementing $\Sigma$ on the fields as
$\tilde{\rho}=a^3\rho
$ and
$\tilde{\bf u}=a\,{\bf u}-\dot{a}\,{\bf x}$,
the equations
of viscosity--free polytropic
hydrodynamical system with $\chi=1$
are  invariant when $a(t)\propto t$ and $\gamma=5/3$.

Curiously, their $\Sigma$
appeared before in cosmology.
The relation is explained as follows.
 In the uniformly expanding newtonian cosmological model \cite{SOU},
the gravitational acceleration has the form
${\bf g}=-(B/a^3){\bf x}$,
where $B$ is a constant related to the scale factor $a(t)$
as $B=-a^2\ddot{a}$.  The Hubble constant is $H=\dot{a}/a$,
and ${\bf g}$ satisfies
$\vec\nabla\cdot{\bf g}=-4\pi G\rho$
(rather than the Einstein equations, as in relativity).
Combining this with $\dot{\bf x}=H{\bf x}$ and $\ddot{\bf x}={\bf g}$ yields
\begin{equation}
    \big(\dot{a}\big)^2=\frac{2B}{a}-K
    \qquad\hbox{and}\qquad
    \rho=\frac{3B}{4\pi Ga^{3}},
\end{equation}
where $K$ is another constant now unrelated to space curvature.
This non-relativistic model is, however,
equivalent to the relativistic
Friedmann universe with constant curvature $K$ \cite{HE}.
The model is also conveniently described \cite{DGH}
by the (``Kaluza--Klein'') $5$-metric
\begin{equation}
    d{\bf x}^2+2dtds-\frac{B\,{\bf x}^{2}}{a^3}dt^2,
    \label{expuniv}
\end{equation}
whose gravitational field equation requires indeed
$\bigtriangleup \big(B{\bf x}^2a(t)^{-3}\big)=8\pi G\rho$ as above.
Now this metric can be conformally mapped to flat space
with metric $d{\tilde{\bf x}}^2+2d\tilde{t}d\tilde{s}$, using
\begin{equation}
    \tilde{\bf x}=\frac{{\bf x}}{a},
    \qquad
    \tilde{t}=\int \frac{dt}{a^{2}},
    \qquad
    \tilde{s}=s+\smallover1/2H{\bf x}^2.
    \label{confflat}
\end{equation}
The (inverse of) (\ref{confflat}) carries the flat-space
hydrodynamical equations
into those valid in the expanding universe.

For the choice of Drury and Mendon\c ca   $B=0$, so that
their expanding metric (\ref{expuniv}) is flat
and has therefore little cosmological interest since then also $\rho=0$.
Ignoring this aspect, we note that the transformation (\ref{confflat}),
which becomes now precisely $\Sigma$ completed with
$s\to s+{\bf x}^2/2t$,  is a conformal transformation of flat space into
itself. The invariance of the Euler equations
under $\Sigma$ follows. This is of course consistent with $\Sigma$
belonging to the ${\rm SL}(2,\IR)$ invariance group of the free system
discussed above. Unfortunately, this symmetry is broken by the viscosity.

Interestingly, the  map $\Sigma$ has also been used
to solve planetary motion when the gravitational constant changes
inversely with time \cite{VIN, DGH}.
It is worth mentionning also that a Friedmann metric
containing a perfect fluid with equation of state
$p=(\gamma-1)\rho$ has also been studied \cite{BARROW}.

\section{Schr\"odinger fields and the Madelung fluid}

Let us conclude with a remark on the  well-known
Schr\"odinger invariance
of the non-linear Schr\"odinger equation
$i\partial_{t}\psi=-\bigtriangleup\psi/2+\lambda\,\vert\psi\vert^{4/d+1}\psi$.
Decomposing the Schr\"odinger field
into module and phase, $\psi=\sqrt{\rho}\,e^{i\phi}$, yields in fact
the hydrodynamical system referred to as the Madelung
fluid \cite{MAD},
\begin{eqnarray}
\partial_t\rho &+ &\vec{\nabla}\cdot(\rho\vec{\nabla}\phi)=0,
\\
\partial_t\phi&+&\frac{1}{2}\vert\vec{\nabla}\phi\vert^2
=-\frac{1}{4\rho}\left[\frac{1}{2}
\frac{\vert\vec{\nabla}\rho\vert^2}{\rho}
-{\Delta\rho}\right]
+\partial_{\rho}U,
\label{madelungequations}
\end{eqnarray}
where $U=\lambda\rho^{(2/d+1)}$.
Eqns $(3.1)$ and $(3.2)$ can be obtained from the
irrotational and viscosity--free Euler equations
choosing the field $\chi$ non-trivially,
\begin{eqnarray}
\chi=\frac{d}{8{\Lambda}\rho^{2/d+1}}\,\left[
\frac{\vert{\vec\nabla}\rho\vert^2}{2\rho}
-\bigtriangleup\rho\right].
\label{chirelation}
\end{eqnarray}
Now, as seen above, the general Euler equations with the standard
polytropic exponent
$\gamma=1+2/d$,
 are
Schr\"{o}dinger invariant whenever $\tilde{\chi}=\chi$.
Using (\ref{dilatation}),
we can show that our $\chi$ transforms precisely in this way.
Therefore, the Madelung equations
 are Schr\"{o}dinger invariant.

In is worth noting that for the membrane potential $\gamma=-1$
one can still choose such a $\chi$. However, owing to the bracketed
term, $\widetilde{\chi}\neq\chi$, so that  the Poincar\'e symmetry is
broken.
The non-relativistic conformal symmetries are also broken, and
we are left with a mere Galilei symmetry.

\kikezd{Note added}. After this paper has been accepted, we became
aware of a paper by Bordemann and Hoppe \cite{BoHo2}, which offers
yet another way to derive the Schr\"odinger invariance.
For simplicity,
we only spell this out in the irrotational case $\theta=0$.
Expressing
$\rho$ from the second equation in (\ref{equationsQ}) and
inserting into the first one yields the
so-called ``Steichen equation'',
which in fact derives from the Lagrangian
\begin{equation}
{\cal L}=\left[\partial_t\phi+
\frac{1}{2}\big(\vec\nabla\phi)^2
\right]^{\frac{\gamma}{\gamma-1}}.
\label{BI}
\end{equation}
Under a non-relativistic
dilatation ${\cal L}$ scales as
${\cal L}\to\delta^{2\gamma/1-\gamma}\,{\cal L}$;
taking into account
 the scaling of the volume element, 
invariance is obtained precisely when $\gamma=1+2/d$.
Note that for the Chaplygin value $\gamma=-1$,
(\ref{BI}) becomes the
Lagrangian used by Jackiw and Polychronakos
\cite{JAC}.

\goodbreak
\kikezd{Acknowledgements}.
The authors are indebted to Professor R. Jackiw for his interest, advice and
for sending them his unpublished remark \cite{JPRIV}, as well as
Professor C. Duval and Dr. O. Ley for discussions.
They acknowledge the {\it Institute for Theoretical Physics}
of E\"{o}tv\"{o}s University (Budapest, Hungary) for hospitality
extended to them.

\goodbreak


\begin{thebibliography}{2}


\bibitem{DM}
{\sc L. O'C. Drury and J. T. Mendon\c ca}, astro-ph/0003385,
to appear in {\em Physics of Plasmas}, Jan. (2001).

\bibitem{NH}
{\sc R. Jackiw}, {\em Phys. Today} {\bf 25},\, 23\,
(1972);
{\sc U. Niederer}, {\em Helv. Phys. Acta} {\bf 45},\,
802\, (1972);
{\sc C. R. Hagen}, {\em Phys. Rev. D} {\bf 5},\, 377\, (1972).

\bibitem{DGH}{\sc
C. Duval, G. Gibbons and P. Horv\'athy},
{\em Phys. Rev.  D} {\bf 43},\, 3907\, (1991).
The non-relativistic
Kaluza-Klein--type framework was  proposed in
{\sc C. Duval, G. Burdet, H. P. K\"{u}nzle and
M. Perrin}, {\em Phys. Rev. D} {\bf 31},\, 1841\, (1985).

\bibitem{RS}{\sc L. O'Raifeartaigh and V. V. Sreedhar},
hep-th/0007199, submitted to {\em Physics of Fluids}.

\bibitem{BoHo}{\sc M.~Bordemann and J.~Hoppe},
{\em Phys. Lett.} {\bf B317},\,315\, (1993);
{\sc A. Jevicki}, {\em Phys. Rev.} {\bf D 57}, R5955 (1998).

\bibitem{JAC}{\sc R. Jackiw and A. P. Polychronakos}, {\em
Comm. Math. Phys.} {\bf 207},\, 107\, (1999); {\sc R. Jackiw},\,
(hep-th/9911235)\,(1999). For a review, see
{\sc R. Jackiw},\, (physics/0010042).

\bibitem{BJ}{\sc D. Bazeia and R. Jackiw}, {\em Ann. Phys.} {\bf
270},\, 146\, (1998);
{\sc D. Bazeia}, {\em Phys. Rev. D} {\bf 59},\, 085007\, (1999).

\bibitem{HH}{\sc M. Hassa\"{\i}ne and P. Horv\'athy}, {\em
Ann. Phys.} {\bf 282},\, 218\, (2000).

\bibitem{JPRIV}
{\sc R. Jackiw}, Private Communication (2000).

\bibitem{JP}
{\sc R. Jackiw and S.-Y. Pi},
{\em Nucl. Phys.} {\bf B} (Proc. Suppl). {\bf 33C}, 104 (1993).

\bibitem{CLE}{\sc H. Lamb}, Hydrodynamics, Cambridge University Press, 1942.


\bibitem{SOU}
{\sc J.-M. Souriau}, in {\em G\'eom\'etrie Symplectique et Physique
Math\'ematique}. Coll. Int. CNRS,  No. {\bf 237},
ed. J.-M. Souriau. Editions du CNRS, p. 59 (1975).

\bibitem{HE}
{\sc S. W. Hawking and G. F. R. Ellis},
 {\em The Large Scale Structure of space-time}. Cambridge  U. P. (1973).

\bibitem{VIN}
{\sc J. P. Vinti},
{\em Mon.  Not. R. Astron. Soc.} {\bf 169}, 417  (1974);
{\sc J. D. Barrow},
{\it ibid}. {\bf 282}, 1397 (1996).

\bibitem{BARROW}
{\sc J. D. Barrow},
{\em The Observatory} {\bf 113}, No 1115, 210 (1993);
{\sc H. C. Rosu},
gr-qc/0003108, {\em Mod. Phys. Lett.} {\bf A 15},  979 (2000).

\bibitem{MAD}{\sc E. Madelung}, {\em Z. Phys} {\bf 40},\, 332\, (1926).

\bibitem{BoHo2}{\sc M.~Bordemann and J.~Hoppe},
{\em Phys. Lett.} {\bf B325},\,359\, (1994).

\end{thebibliography}
\end{document}